\documentclass[doublecol]{epl2} 
\usepackage[T1]{fontenc}
\usepackage[latin1]{inputenc}
\usepackage{graphicx}
\usepackage{amsmath}
\usepackage{amstext}
\usepackage{amssymb}

\begin{document}

\title{Loss of synchronization in complex neuronal networks with delay}
\shorttitle{Loss of synchronization in complex neuronal networks with delay}

\author{Judith Lehnert\inst{1} \and Thomas Dahms\inst{1} \and Philipp
  H\"{o}vel\inst{1,2} \and Eckehard Sch\"{o}ll\inst{1}\thanks{E-mail: \email{schoell@physik.tu-berlin.de}}}

\shortauthor{J. Lehnert \etal}

\institute{
\inst{1} Institut f{\"u}r Theoretische Physik, Technische
  Universit{\"a}t Berlin, Hardenbergstra{\ss}e 36, 10623 Berlin,
  Germany\\
\inst{2} Bernstein Center for Computational Neuroscience Berlin,
Philippstra{\ss}e 13, Haus 2, 10115 Berlin, Germany}

%\keywords{synchronization, networks, inhibition, neural oscillators}
% Synchronization, nonlinear dynamics, 05.45.Xt
% neural networks, 87.85.dq
% Delay equations, in function theory, 02.30.Ks
% Complex systems, 89.75.-k

\abstract{
  We investigate the stability of synchronization in networks of
  delay-coupled excitable neural oscillators. On the basis of the
  master stability function formalism, we demonstrate that
synchronization is
  always stable for excitatory coupling independently of the delay and
  coupling strength. Superimposing inhibitory links randomly on top of
  a regular ring of  excitatory coupling, which yields 
  a small-world-like network topology,
we find a phase transition to desynchronization as the probability
of inhibitory links exceeds a critical value.
We explore the scaling of the
critical value in dependence on network properties.
Compared to random networks, we find that small-world topologies are
more susceptible to desynchronization via inhibition.
}

\maketitle
\section{Introduction}
Studies of complex networks have sparked tremendous scientific
activities in many research areas and the analysis of network
topologies in real-world systems has become a field of large
interest. For instance, there is evidence that neuronal networks on
the level of single neurons coupled through synapses or gap junctions,
as well as on the level of cortex areas and their pathways exhibit the
small-world (SW) properties \cite{WAT98,SPO00SHE02SPO06HON07SPO07}.
The high clustering coefficient of the SW networks enhances local
communication efficiency, while the small shortest path length enables
efficient global communication \cite{LAT01}. Thus, the SW architecture
is optimal for processing and transmission of signals within and
between brain areas. However, the synchronizability of small-world
networks depends in a delicate way upon the network topology \cite{NIS03}. Next to this structural aspect, inhibition plays
a prominent role in many neural processes \cite{HAI06}.  Without an
inhibitory mechanism, excitation in a compound system would not decay,
but spread through the whole network, finally leading to persistent
spiking of all neurons. Thus, encoding and processing of information
would be impossible.

In this Letter, we combine both fundamental aspects -- inhibition and
SW property -- in order to emphasize the important interplay of
excitation and inhibition in complex networks. We start with a regular
ring network that consists of purely excitatory links with
delay. Thus, it exhibits strong and stable synchronization.  Depending
on the initial conditions, both isochronous and cluster
synchronization are possible implying multistability. Additional
inhibitory connections, which we include in a SW-like manner
% \cite{WAT98,MON99,NEW99b}
\cite{WAT98,MON99}, result in a loss of synchronization.  A similar
transition was reported for phase oscillators in Ref.~\cite{TOE10} for
unidirectional rings, but the effect of inhibition upon excitable
systems could not be treated by that model.  For the node dynamics, we
consider a generic model to demonstrate the fundamental relevance and
importance of our findings in the field of neuroscience. In this area,
synchronization can be related to cognitive capacities \cite{SIN99b}
as well as to pathological conditions, e.g., epilepsy \cite{UHL06}. A
better understanding of the loss of synchronization will eventually
lead to future therapeutic treatments \cite{HAU07}.

Throughout this Letter, we consider a network of $N$ delay-coupled
FitzHugh-Nagumo (FHN) oscillators. The FHN system describes neuronal
dynamics by a two-variable model
% \cite{FIT61,NAG62}
\cite{FIT61}. Because of its simplicity it can be considered as a
paradigmatic model of excitable systems, which also occur in several
other natural contexts ranging from cardiovascular tissues to the
climate system
% \cite{MUR93,MIK94,KEE98,IZH00a,KOC99,WUE02,GAN02}
\cite{MUR93,IZH00a}. Here, the network
dynamics is described by
\begin{subequations} \label{fhn:network}
\begin{eqnarray}
\epsilon \dot{u}_i &=& u_i - \frac{u_i^3}{3} - v_i+ C\sum_{j=1}^N G_{ij}[u_j(t-\tau)-u_i]\\
\dot{v}_i &=& u_i + a,
\end{eqnarray}
\end{subequations}
where $u_i$ and $v_i$ denote the activator and inhibitor variables of
the nodes $i=1,\ldots,N$, respectively. The parameter $a$ determines
the threshold of excitability. A single FHN oscillator is excitable
for $a>1$ and exhibits self-sustained periodic firing beyond the Hopf
bifurcation at $a=1$. Here, we will focus on the excitable regime with
$a=1.3$. The time-scale parameter $\epsilon$ is chosen as $\epsilon =
0.01$.  $C$ is the coupling strength.  $\mathbf {G}=\{G_{ij}\}$,
$i,j=1,\ldots,N$, denotes the coupling matrix that determines the
topology of the network. In the following we will assume unity row sum
of $\bf G$.  This ensures that each neuron receives the same input if
the network is synchronized. The delay time $\tau$ takes into account
the finite propagation speed of an action potential.  We investigate
complete synchronization with $(u_i(t),v_i(t))=(u_s(t),v_s(t))\equiv
\mathbf{x}_s(t)$ for $i=1,\dots,N$, which is also known as zero-lag or
isochronous synchronization.  This state is a solution of
qs.~\eqref{fhn:network} and reduces the system's dynamics to
\begin{equation}
  \label{eq:sync}
  \dot{\mathbf{x}}_s = \mathbf{F}(\mathbf{x}_s) + C \mathbf{H}
\left[\mathbf{x}_s(t-\tau)-\mathbf{x}_s(t)\right]
\end{equation}
with $\mathbf{F}(\mathbf{x})=\left(\begin{smallmatrix}[u - u^3/3
    -v]/\epsilon\\u+a\end{smallmatrix}\right)$ % . The matrix
and the matrix $\mathbf{H} =\left(\begin{smallmatrix}1/\epsilon & 0\\0 &
    0\end{smallmatrix}\right)$.  The $2(N-1)$ constraints of complete
synchronization define a two-dimensional synchronization manifold (SM)
in the $2N$-dimensional phase space.

\begin{figure}[t]
  \centering
  \includegraphics[width=\columnwidth]{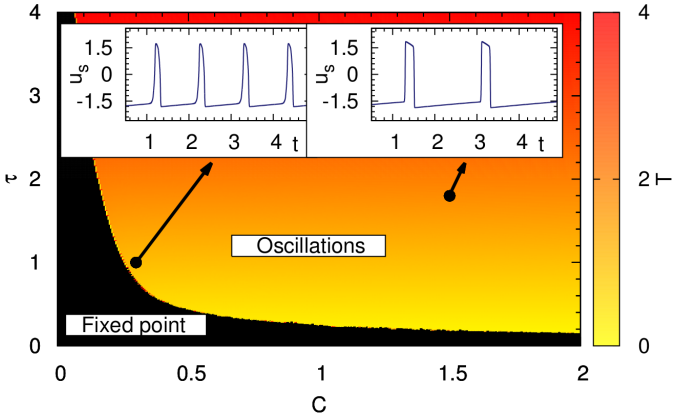}
  \caption{(Color online) Dynamics in the synchronization manifold in
    dependence on the coupling strength $C$ and delay $\tau$. The
    gray scale (color code) indicates the period of spiking
    oscillations $T$, the black region corresponds to fixed-point
    dynamics. Left and right insets show time series of the activator
    $u_s$ for ($C=0.3$, $\tau=1$) and ($C=1.5$, $\tau=1.8$),
    respectively. Parameters: $a=1.3$, $\epsilon=0.01$.}
  \label{fig:syncdynamics}
\end{figure}

As we operate in the excitable regime, the dynamics on this SM, in
particular the period, will depend on the choice of the coupling
parameters $C$ and $\tau$ as depicted in fig.~\ref{fig:syncdynamics}.
The grayscale (color code) corresponds to the period $T$ of the
oscillations on the SM, which we find to follow $T=\tau+\delta$ with
$\delta \ll \tau$ accounting for a short activation time
% \cite{SCH08,DAH08c}
\cite{SCH08}. For small coupling strength $C$ the incoming signal is
not sufficient to trigger oscillations (black region). For small delay
times $\tau$ consecutive spikes run into the refractory phase of the
previous one, which prevents oscillations as well. From here on we
consider $C$ and $\tau$ sufficiently large such that the coupling
induces oscillations.

\section{Stability analysis} 
In the following we address the question
whether the oscillatory solution on this manifold is transversely
stable. The master stability function (MSF) \cite{PEC98} allows us to
quantify this transversal stability. It can be calculated as largest
Lyapunov exponent $\Lambda$ from eq.~\eqref{fhn:network} 
linearized around eq.~\eqref{eq:sync}:
\begin{eqnarray} \label{eq:msf}
\dot{\boldsymbol \zeta}(t)= \left[ D\mathbf{F}(\mathbf{x}_{s}(t)) -C
  \mathbf{H} \right] \boldsymbol \zeta(t) + (\alpha+i\beta) \textbf{H}
\boldsymbol \zeta(t-\tau).
\end{eqnarray}
Here, $D\mathbf{F}$ denotes the Jacobian of $\mathbf{F}$.  The idea of
the MSF is to calculate the stability of a synchronized solution for
an arbitrary topology matrix $\mathbf{G}$. For this purpose, the
parameter $\alpha + i\beta$ represents a continuous parametrization of
$\{C \nu_i\}$, where $\nu_i $, $i=1,\ldots,N$, are the eigenvalues of
$\mathbf G$. In the same sense, the vector $\boldsymbol \zeta$ is a
generalization of the variational vectors transformed to the
corresponding eigensystem.  In the ($\alpha$,$\beta$)-plane the MSF
typically gives rise to regions with negative $\Lambda$.  If all
rescaled transversal eigenvalues $\{C \nu_i\}$ of a given network are
located within this stable region, perturbations from the SM will
decay exponentially and the synchronized dynamics will be stable. Due
to the unity row sum condition, $\mathbf G$ will always have one
eigenvalue $\nu_1=1$. This longitudinal eigenvalue is associated with
perturbations within the SM and is not relevant for the stability of
synchronization. $\Lambda(C\nu_1)$ determines the type of dynamics on
the SM.  For periodic dynamics in the SM, as in the present case, we
have $\Lambda(C\nu_1)=0$.

\begin{figure}[t]
  \centering
  \includegraphics[width=1.0\linewidth]{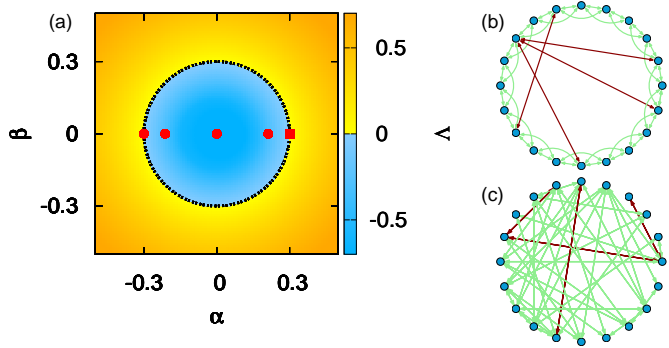}
  \caption{(Color online) (a) Master stability function for a network
    of FHN systems given by eqs.~\eqref{fhn:network}. Dotted curve:
    $S((0,0),C)$. Red circles (square): Rescaled transversal
    (longitudinal) eigenvalues $C \nu_i$ of a bidirectionally coupled
    ring with $N=8$ nodes. Parameters: $a=1.3$, $\epsilon=0.01$,
    $C=0.3$, $\tau=1$.  (b) Scheme of a bidirectional regular network
    ($N=20$, $k=2$), and (c) a random network ($N=20$, fixed number of
    links $kN$) with excitatory coupling (gray (green) arrows) on
    which inhibitory links (black (red) arrows) are superimposed.}
  \label{fig:msf}
\end{figure}

Figure~\ref{fig:msf} depicts the MSF for the network of FHN
oscillators given by eqs.~\eqref{fhn:network}. Dark (blue) colors mark
the stable region. As an illustration the rescaled eigenvalues
$\{C\nu_i\}$ of a bidirectionally coupled ring ($N=8$) are shown as
red symbols. The corresponding coupling matrix is given by
$G_{i,i+1\bmod{N}}=G_{i,i-1\bmod{N}}=1$ ($i=1,\ldots,N$) and zero
otherwise. The rescaled longitudinal eigenvalue $C\nu_1=C$ is depicted
by a black (red) square. All rescaled transversal eigenvalues (black
(red) circles) lie inside the stable region indicating that the
synchronization of the bidirectionally coupled ring is stable.

\section{Shape of stability region} 
The MSF must be calculated for each combination of $C$ and
$\tau$. Although different $C$ and $\tau$ lead to quantitatively
different Lyapunov exponents $\Lambda$, the shape of the stable
regions remains qualitatively very similar. In particular, it is in
very good approximation given by the circle $S((0,0),C)$ with center
at the origin and radius $C$ (dotted circle in fig.~\ref{fig:msf})
independent of the specific values of $C$ and $\tau$. The rotational
symmetry has recently been proved generally for large $\tau$
\cite{FLU10b}. Only for small $\tau$ and $C$ the stable region is
slightly larger than the circle and shows a bulge around $\alpha=-C$,
$\beta=0$ \cite{remark}.  The positive $\alpha$-axis is always
intersected at $\alpha=C$, which corresponds to $\Lambda(C\nu_1)=0$ as
discussed above.  For any choice of $\tau$ and $C$ that leads to
periodic dynamics on the SM, the circle $S((0,0),C)$ serves as a lower
bound for the stability boundary. See the appendix for an
  analytic derivation of this circle $S((0,0),C)$ in the limit of
  large coupling strength and as a lower bound for all coupling
  strengths.  We conclude that the stability of the synchronized
periodic dynamics, if such a solution exists, depends only on the
topology and neither on the coupling strength nor on the delay time.

\section{Excitatory coupling}
 For excitatory coupling, i.e.,
$G_{ij}\geq 0$, all eigenvalues of $\mathbf{G}$ are located inside the
stable region.  Using Gershgorin's circle theorem
% \cite{EAR03,CHO09}
\cite{EAR03}, which gives an upper bound of the eigenvalues, and the
constant row sum assumption, all Gershgorin circles ($i=1,\dots,N$),
centered at $G_{ii}$ with radius $\sum_{j\neq i}G_{ij} = 1-G_{ii}$
because of the unity row sum, lie inside the unit circle. Thus all
rescaled eigenvalues $\{C\nu_i\}$, $i=1,\ldots,N$, are located inside
$S((0,0),C)$, i.e., inside the stable region. Networks with purely
excitatory coupling will always exhibit stable synchronization.

\section{Inhibitory coupling} 
As a consequence of this result,
desynchronization can only be achieved by introducing negative entries
in the coupling matrix {\bf G}, i.e., inhibitory coupling between
neurons. This inhibition is a crucial feature in neural
processes, %\cite{KAN96,POE01,JON07,OKU09}
e.g., to overcome unwanted synchronization associated with
pathological states.

\begin{figure}[t]
  \centering
  \includegraphics[width=\linewidth]{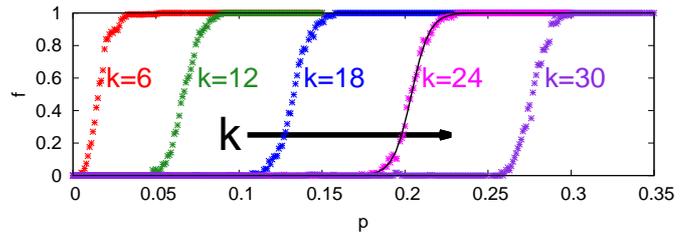}
  \caption{(Color online) Fraction of desynchronized networks $f$ vs
    the probability of additional inhibitory links $p$ for
    $N=100$. $k$ varies from 6 to 30. Thin black curve: Example fit to
    $f(p)$ ($p_c=0.20387$, $b=186$) for $k=24$.  Number of
    realizations: 500 for each value of $k$. Parameters as in
    fig.~\ref{fig:syncdynamics}.}
  \label{fig:phasetrans}
\end{figure}

Particularly, we consider the following variation of the
Watts-Strogatz SW network
% \cite{WAT98,MON99,NEW99b}
\cite{WAT98,MON99}: (i) Start with a one-dimensional ring of $N$
nodes, where every node is connected by excitatory links to its $k$
neighbors on either side. (ii) For each of the $kN$ links of the
network add an inhibitory link with probability $p$ connecting two
randomly chosen nodes.  (iii) Do not allow self-coupling or more than
one link between any pair of nodes.  (iv) Normalize the entries of the
coupling matrix $\textbf{G}$ by dividing each row by the absolute
value of its row sum. In the case that the row sum of the $i$th row is
negative we set $G_{ii}=2$ to ensure unity row
sum. Figure~\ref{fig:msf}(b) illustrates such a SW network for $N=20$
and $k=2$, where gray (green) and black (red) arrows indicate
excitatory and inhibitory coupling, respectively.  For each
realization of such a network, we determine the stability of
synchronization by checking whether the full eigenspectrum ${C\nu_i}$
of the coupling matrix is contained in the stable region
$S((0,0),C)$. Hereby we compute the fraction $f$ of desynchronized
networks.  Figure~\ref{fig:phasetrans} shows $f$ as a function of $p$
for different coupling ranges $k$.
This Figure is virtually identical for all delay times, that is, for all 
parameters within the color shaded area of Fig.1. To obtain this Figure we 
made use of the circular shape of the stable region of the master stability 
function. Only for very small delays or coupling strength, the
stable region is slightly larger and thus the shape of the curves
shown in Figure~\ref{fig:phasetrans} might be shifted slightly to larger values of $p$.

\begin{figure}[t]
  \centering
 \includegraphics[width=0.48\columnwidth]
 {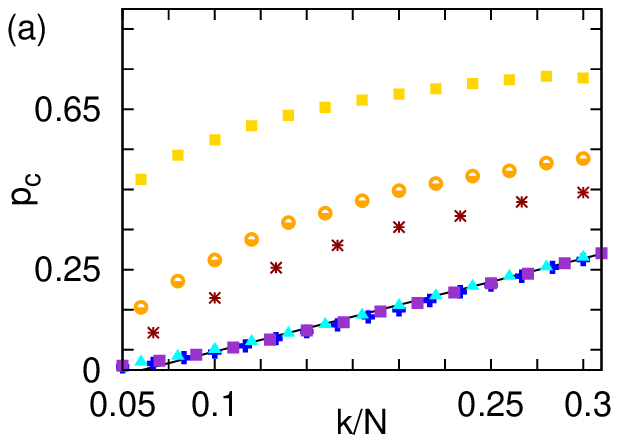}
 \includegraphics[width=0.48\columnwidth]
 {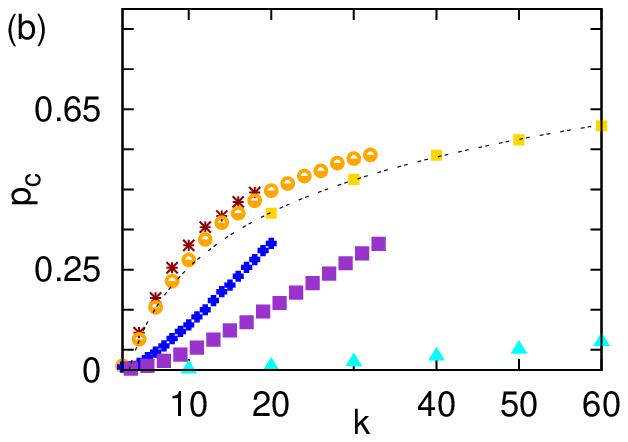}
 \includegraphics[width=0.48\columnwidth]
 {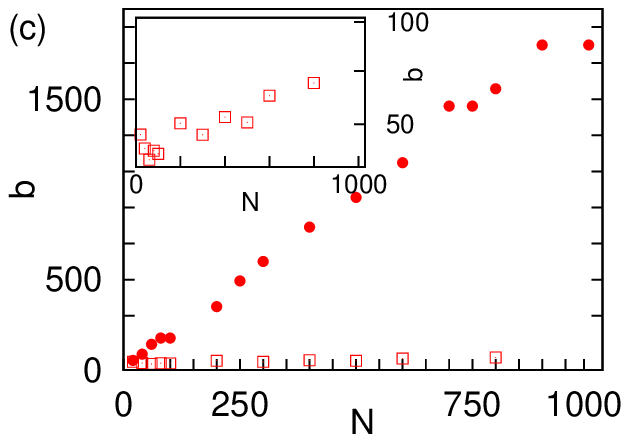}
\includegraphics[width=0.48\columnwidth]
 {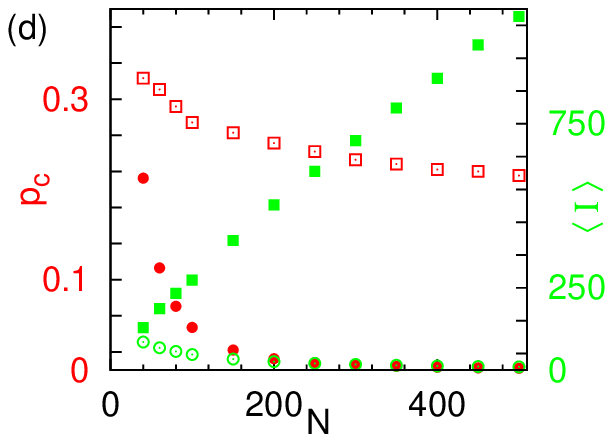}
 \caption{(Color online) Critical value $p_c$ for different network
   sizes, (a) in dependence on $k/N$, (b) in dependence on $k$: SW
   networks $N=60$ (dark (blue) circles), $N=100$ (dark (purple)
   squares), and $N=500$ (turquoise triangles). Random networks $N=60$
   (black (red) crosses), $N=100$ (light (orange) circles), and
   $N=500$ (lightgray (yellow) squares).  (c) Steepness $b$ for SW
   (circles) and random (squares) networks vs. $N$ for constant
   $k/N=0.1$. Inset: Blow-up for random networks. (d) $p_c$ vs. $N$
   for $k=10$ for a SW (black (red) filled circles) and a random
   (black (red) empty squares) network. Number of inhibitory links
   $\langle I \rangle$ vs. $N$ for constant $k$ for a SW (gray (green)
   empty circles) and a random (gray (green) filled squares) network.
   Number of realizations: 500.}
  \label{fig:pcrit}
\end{figure}

For fixed $k$ a steep transition between synchronization and
desynchronization takes place as $p$ approaches a critical value
$p_c$. This critical value $p_c$ and the steepness $b$ of the
transition can be fitted with a sigmoidal function
$f(p)=1/[e^{-b(p-p_c)}+1]$. Figure~\ref{fig:pcrit}(a) depicts the
critical probability $p_c $ for $f(p_c)=0.5$ in dependence on $k/N$
for different network sizes.  It can be seen that for SW networks
$p_c$ follows a linear relation $p_c(k/N) = 1.16 k/N -0.07$
independently of the network size $N$. Figure~\ref{fig:pcrit}(c) shows
the steepness $b$ as a function of the network size demonstrating that
the transition becomes increasingly sharp as $N$ increases. This
indicates a first-order nonequililibrium phase transition in the 
thermodynamic limit \cite{SCH87}.

To verify whether this phase transition and especially its
independence of the network size is common in networks with inhibitory
links or unique to the SW structure, we construct a different network
for comparison: The regular excitatory network is replaced by a random
network with fixed number of excitatory links $kN$ equivalently to the
regular network used before, as shown in fig.~\ref{fig:msf}(c). We
only consider realizations where this underlying excitatory network is
fully connected. The construction of the inhibitory links then follows
steps (ii-iv) as above. We find that a phase transition to
desynchronization still occurs with critical probabilities of
inhibitory links $p_c$ as depicted in fig.~\ref{fig:pcrit}(a) by black
(red) crosses, gray (orange) circles, and lightgray (yellow) squares
for $N=60$, $100$, and $500$, respectively. We observe, however, that
the values of $p_c$ are higher, i.e., the random network can tolerate
more inhibitory links than the SW network before
desynchronizing. Furthermore, the function $p_c(k/N)$ is no longer
independent of the network size $N$.  Instead, $p_c$ is a function of
$\log(k)$ as can be seen in fig.~\ref{fig:pcrit}(b), where $p_c$ is
plotted in dependence on $k$ for the different network sizes $N$.
Figure~\ref{fig:pcrit}(d) depicts $p_c$ in dependence on $N$ for
constant $k=10$ for a random (black (red) empty squares) and for a SW
network (red (gray) circles).  For random networks, $p_c$ is
independent of $N$ for sufficiently large $N$, while for SW networks
it approaches zero. Recall that $p_c$ is the mean value of the ratio
of inhibitory to excitatory links. Thus, we conclude that in SW
networks with increasing network size $N$ but same local structure
(constant $k$) an infinitesimally small ratio of inhibition to
excitation is needed to prevent synchronization, while in a random
network even for very large networks only a non-vanishing ratio
impairs synchronization. We find in a SW network with constant $k$
that the mean value of the number of inhibitory links $\langle I
\rangle:= p_ckN$ causing desynchronization scales as $\langle I
\rangle= 1.16k^2-0.07kN$ for small $N$ and approaches zero for large
$N$ (see fig.~\ref{fig:pcrit}(d) green (gray) empty circles). In
contrast, in a random network $\langle I \rangle$ is proportional to
$N$ (see fig.~\ref{fig:pcrit}(d) green (gray) filled squares), i.e.,
an increasing number of inhibitory links is needed.  This difference
to SW networks can be understood in an intuitive way: In a SW, any
added inhibitory link is part of a shortest path for many pairs of
nodes, as it shortens the mean path length considerably with respect
to the underlying regular ring. In the random network, however, where
the mean path length is relatively low even without added shortcuts,
only few node pairs will gain shorter paths by adding inhibitory
links. Considering the dynamics on a network, perturbations
  from the synchronized state spread along the shortest paths first,
  changing the response of the receiving node, and information flow
  along longer paths will reach the receiving node only at a later
  time and will not influence the change of the initial response. In
  conclusion, if a large fraction of the inhibitory links is part of
  the shortest paths - like in the small-world topology superimposed
  to a regular ring - these inhibitory shortcuts become dominant.

\section{Conclusions} 
We have shown how the interplay of excitatory and inhibitory couplings
leads to desynchronization in networks of neural oscillators. The
desynchronization is achieved via a phase transition from a completely
synchronized state. This can be seen as a first step towards an
understanding of the robustness of different states of synchrony,
e.g., cluster synchronization, in arbitrary networks with weighted
links or distributed delays. Note that for appropriate network
topologies the framework of the MSF presented above can indeed be
extended to cluster synchronization where the oscillators synchronize
in $M$ clusters with a constant phase lag $2\pi/M$ between subsequent
clusters% \cite{SOR07,KAN11a}
\cite{SOR07}.  The corresponding SM is $2M$ dimensional. Hence, $M$
longitudinal eigenvalues exist. The MSF, however, is again very well
approximated by the circle $S((0,0),C)$ and thus, we observe
multistability between zero-lag and cluster synchronization.

Excitable systems can be classified into type-I and type-II
excitability~\cite{HOD48,IZH00a}. In addition to the generic type-II
FitzHugh-Nagumo model used in this paper, we have considered the
normal form of a {\em saddle-node bifurcation on an invariant circle}
(SNIC) as a generic model of type-I excitability~\cite{HU93a}. For
sufficiently large delay times and coupling strength the MSF is again
given by the circle $S((0,0),C)$ implying that the previously obtained
results persist. In particular, the same phase transition occurs. This
indicates that the phenomena observed here are generic for any
excitable system.

\section{Appendix: Analytic approximation of the stability region}

The numerical calculation of the master stability function has shown that
$S((0,0),C)$ is a lower bound and a very good approximation of the
stable region for all $\tau$ and $C$. As $\tau$ and $C$ increase the
approximation becomes even better. A Taylor expansion as done in
Ref.~\cite{JUS97} for the investigation of time-delayed feedback control of an
unstable periodic orbit gives analytic insight in the problem.  This
analysis is very general and does not use the specific form of the
local dynamics in terms of the FHN model. It only assumes that the
synchronized dynamics is oscillatory with period $T$.  Using a Floquet
ansatz ${\boldsymbol \zeta}=e^{(\Lambda+i\Omega)t} \mathbf{Q} (t)$ with the
periodic function $ \mathbf{Q} (t)= \mathbf{Q} (t+T)$ in
Eq.~\eqref{eq:msf} yields
\begin{align} 
  &(\Lambda+i\Omega) \mathbf{Q} (t)+\dot{ \mathbf{Q}} (t) \\
  &\quad =(D {\bf F} -C {\bf H}) \mathbf{Q} (t)
  +(\alpha +i \beta)e^{-(\Lambda+i\Omega)\tau}{\bf H} \mathbf{Q}
  (t-\tau). \nonumber
\end{align}
$\Lambda+i\Omega$ is the Floquet exponent, whose real part coincides
with the Lyapunov exponent in the case of a periodic orbit.
 
Assume $T=\tau$. In the case of the FHN system this is an
approximation since the period of the oscillations differs by a small
activation time $\delta \ll \tau$ from the delay time $\tau$ following
$T=\tau+\delta$. Then $\mathbf{Q}(t-\tau)$ can be substituted by
$\mathbf{Q}(t)$:
\begin{align} \label{eiglsa}
  &(\Lambda+i\Omega) \mathbf{Q} (t)+\dot{ \mathbf{Q}} (t) \\
  &\quad =(D {\bf F} ) \mathbf{Q} (t) + \underbrace{[-C+(\alpha+i
    \beta)e^{-(\Lambda+i\Omega)\tau}]}_{\kappa}{\bf H} \mathbf{Q}
  (t). \nonumber
\end{align}
We expand the solution $\Gamma(\kappa)=\Lambda + i\Omega$ of the
eigenvalue problem defined by Eq.~\eqref{eiglsa} in a Taylor
approximation:
\begin{eqnarray}
\Gamma(\kappa)=\Gamma(0)+\Gamma'(0)\kappa+O(\kappa^2).
\end{eqnarray}
Using $\Gamma(0) \equiv \lambda+i\omega$
and $\Gamma'(0)\equiv \chi'+i\chi''$ we obtain
\begin{eqnarray} \label{floquetapprox_allg}
\Lambda+i\Omega=\lambda+i\omega+(\chi'+i\chi'')[-C+(\alpha+i \beta)e^{-(\Lambda+i\Omega)\tau}].
\end{eqnarray}
Note that $\kappa=0$ if $(\alpha,\beta)=(C,0)$ corresponding to the dynamics
within the synchronization manifold. Thus the first term in the Taylor
approximation corresponds to the Goldstone mode, i.e.,
$\lambda+i\omega=0$ for
$(\alpha,\beta)=(C,0)$. Equation~\eqref{floquetapprox_allg} then
becomes
\begin{eqnarray} \label{eqlam}
\Lambda+i\Omega=\chi' [-C+(\alpha+i \beta)e^{-(\Lambda+i\Omega)\tau}].
\end{eqnarray}
Here, we assume $\chi''=0$. Separating Eq.~\eqref{eqlam} into real and
imaginary part leaves us with
\begin{eqnarray} \label{floquetapprox}
\Lambda &=& \chi'\{-C+e^{-\Lambda \tau}[\alpha \cos(\Omega \tau)+\beta \sin(\Omega \tau)]\}, \nonumber \\
\Omega &=& \chi' e^{-\Lambda \tau}[-\alpha \sin(\Omega \tau)+\beta \cos(\Omega \tau)].
\end{eqnarray}
Equation~\eqref{floquetapprox} can be solved numerically yielding the
circular stability region. On the border of the stability the real
part of the Floquet exponent vanishes. Using $\Lambda=0$ in
Eq.~\eqref{floquetapprox} yields after algebraic manipulations:
\begin{eqnarray} \label{eq:bound3}
\alpha_b&=& \frac{-\Omega \sin(\Omega \tau)}{ \chi'} + C \cos(\Omega \tau), \nonumber\\
\beta_b&=& \frac{\Omega \cos(\Omega \tau)}{ \chi'} + C \sin(\Omega \tau),
\end{eqnarray} 
where $\alpha_b$ and $\beta_b$ denote the values of $\alpha$ and
$\beta$, respectively, on the bounder of stability.  Finally we
obtain
\begin{eqnarray} \label{r_eq:bound3}
\alpha_b^2+\beta_b^2= C^2+ \frac{\Omega^2}{\chi'^2}.
\end{eqnarray} 
Obviously $\alpha_b^2+\beta_b^2> C^2$ holds, demonstrating that
$S((0,0),C)$ is a lower bound for the stable region. For large $C$ the
term $C^2$ on the right hand side dominates. Thus, the boundary of
stability is very well approximated by $S((0,0),C)$ for large coupling
strength.

\begin{acknowledgments}
 This work was supported by DFG in the framework of SFB 910. PH
 acknowledges support by the BMBF under the grant no. 01GQ1001B
 (\textit{F\"orderkennzeichen}). 
\end{acknowledgments}
%\bibliographystyle{prwithtitle}
%\bibliographystyle{prsty-fullauthor}
%\bibliography{ref}

\begin{thebibliography}{10}
\bibitem{remark} E.g., for $C=0.3$ and $\tau=1$ the radius is 0.303 in
  the direction of the negative $\alpha$-axis.
\end{thebibliography}

\end{document}